\documentclass[article,nojss]{jss}

\usepackage{orcidlink,thumbpdf,lmodern}
\usepackage[american]{babel}
\usepackage{amsmath,amssymb}
\usepackage[capitalise,noabbrev]{cleveref}
\usepackage{xfrac}
\usepackage{algorithm,algpseudocode}

\creflabelformat{equation}{#2\textup{#1}#3}

\newcommand{\parenth}[1]{\left(#1\right)}
\newcommand{\bracket}[1]{\left[#1\right]}
\newcommand{\braces}[1]{\left\{#1\right\}}
\newcommand{\proba}[1]{p\parenth{#1}}
\renewcommand{\E}[1]{\mathbb{E}\parenth{#1}}
\renewcommand{\vec}[1]{\mathbf{#1}}
\newcommand{\mat}[1]{\mathbf{#1}}
\newcommand{\pred}{\mathcal{P}}
\newcommand{\BigO}{\mathcal{O}}
\DeclareMathOperator{\cov}{cov}
\newcommand{\floor}[1]{\left\lfloor#1\right\rfloor}
\newcommand{\imag}{\mathrm{i}}
\newcommand{\dist}{\sim}
\newcommand{\abs}[1]{\left\vert#1\right\vert}

\author{Till Hoffmann~\orcidlink{0000-0003-4403-0722}\\Harvard T.H. Chan\\School of Public Health
   \And Jukka-Pekka Onnela~\orcidlink{0000-0001-6613-8668}\\Harvard T.H. Chan\\School of Public Health}
\Plainauthor{Till Hoffmann, Jukka-Pekka Onnela}

\title{\pkg{gptools}: Scalable Gaussian Process Inference with \proglang{Stan}}
\Plaintitle{gptools: Scalable Gaussian Process Inference with Stan}
\Shorttitle{gptools: Scalable Gaussian Processes}

\Plainkeywords{Gaussian process, Fourier transform, sparse approximation, Stan, Python, R}

\Abstract{
    Gaussian processes (GPs) are sophisticated distributions to model functional data. Whilst theoretically appealing, they are computationally cumbersome except for small datasets. We implement two methods for scaling GP inference in \proglang{Stan}: First, a general sparse approximation using a directed acyclic dependency graph; second, a fast, exact method for regularly spaced data modeled by GPs with stationary kernels using the fast Fourier transform. Based on benchmark experiments, we offer guidance for practitioners to decide between different methods and parameterizations. We consider two real-world examples to illustrate the package. The implementation follows \proglang{Stan}'s design and exposes performant inference through a familiar interface. Full posterior inference for ten thousand data points is feasible on a laptop in less than 20~seconds. Details on how to get started using the popular interfaces \pkg{cmdstanpy} for \proglang{Python} and \pkg{cmdstanr} for \proglang{R} are provided.
}

\Address{
  Till Hoffmann, Jukka-Pekka Onnela\\
  Harvard T.H. Chan School of Public Health\\
  E-mail: \email{thoffmann@hsph.harvard.edu}, \email{onnela@hsph.harvard.edu}\\
  URL: \url{https://tillahoffmann.github.io/}, \url{https://www.hsph.harvard.edu/onnela-lab/}
}

\begin{document}

\section{Introduction}
\label{sec:introduction}

Gaussian processes (GPs) are flexible non-parametric models for functions with applications in time series analysis \citep{Roberts2013}, geospatial statistics \citep{Krige1951}, robotics \citep{Deisenroth2015}, and beyond. More formally, a GP is a distribution over functions $f\parenth{x}$ such that any finite set of $n$ values $\vec{f}=f\parenth{\vec{x}}$ evaluated at $\vec{x}=\braces{x_1,\ldots,x_n}$ follows a multivariate normal distribution \citep{Rasmussen2006}. The distribution is thus fully specified by its mean $\mu\parenth{x}=\E{f\parenth{x}}$ and covariance kernel $k\parenth{x,x'}=\cov\parenth{f\parenth{x},f\parenth{x'}}$. A range of problem-specific kernels have been developed, such as squared exponential and Mat\'ern kernels to model local correlations and sinusoidal kernels to capture periodic signals \citep[Chapter~2]{Duvenaud2014}. In general, evaluating the likelihood requires inverting the covariance matrix $\mat{K}$ obtained by evaluating the kernel $k$ at all pairs of observed locations $\vec{x}$. Unfortunately, the computational cost of inverting $\mat{K}$ scales as $\BigO\parenth{n^3}$, making GPs prohibitively expensive save for relatively small datasets.

Diverse schemes have been developed to approximate the likelihood for larger datasets, such as low-rank approximations of the covariance matrix \citep{Hensman2013}, nearest-neighbor approximations \citep{Wu2022}, and Fourier methods \citep{Hensman2017,Greengard2021}. Numerous packages provide implementations in different programming languages, but most of them focus exclusively on GPs which makes it difficult to integrate GPs in larger models. For example, \pkg{GPflow} \citep{Matthews2017} is a library for the \proglang{Python} \citep{PythonLanguage} machine learning framework \pkg{TensorFlow} \citep{Abadi2016}. It implements many common likelihood functions, e.g., Poisson for count data, Bernoulli for classification, and Student-t for robust regression, but hierarchical models cannot be easily constructed. \pkg{GPyTorch} \citep{Gardner2018} offers similar functionality for \pkg{PyTorch} \citep{Paszke2019}; this implementation can be used as part of the general purpose probabilistic programming framework \pkg{Pyro} \citep{Bingham2019}. Likelihood gradients are readily available through the underlying machine learning frameworks, and both packages implement automatic differentiation variational inference (ADVI) \citep{Kucukelbir2017}.
\pkg{GPy} \citep{GPy} and \pkg{george} \citep{Ambikasaran2015} (both implemented in \proglang{Python}) as well as \pkg{GaussianProcesses.jl} \citep{Fairbrother2022} for \proglang{Julia} \citep{Bezanson2017JuliaLang}, \pkg{mlegp} \citep{Dancik2008} and \pkg{spBayes} \citep{Finley2007SpBayes} (both for \proglang{R} \citep{Rlanguage}), and \pkg{GPML} \citep{Rasmussen2010} for \proglang{MATLAB} \citep{MatlabLanguage} implement GP inference from the ground up giving maximum flexibility. However, this approach requires error-prone manual implementation of likelihoods, their gradients, and dedicated sampling algorithms, such as Gibbs samplers. Further, users need to become intimately familiar with the codebase to develop custom models. While sampling the posterior using Hamiltonian Monte Carlo (HMC) is supported by \pkg{GPy}, \pkg{GaussianProcesses.jl}, and \pkg{GPML}, the packages implement a standard leap-frog HMC sampler, requiring extensive hyperparameter tuning for efficient sampling; \pkg{mlegp} only supports maximum-likelihood estimation of parameters. \pkg{spBayes} obtains samples efficiently using a Gibbs sampler by first marginalizing with respect to the latent Gaussian process, and \proglang{R}-\pkg{INLA} \citep{GomezRubio2020Inla} offers performant inference using the integrated nested Laplace approximation. However, both remain limited by the computational cost of inverting the covariance kernel.
Building on \proglang{Stan} \citep{Carpenter2017}, \pkg{brms} \citep{Burkner2017} offers a high-level \proglang{R} interface for fitting Bayesian models, including GPs, but it currently only supports squared exponential covariance kernels. \pkg{brms} implements a basis-function approximation to apply GPs to larger datasets \citep{Riutort-Mayol2022} which we discuss further in \cref{sec:discussion}.

Despite its popularity, a library for scalable GP inference is lacking for the probabilistic programming framework \proglang{Stan} \citep{Carpenter2017}. Here, we discuss the implementation of two scalable approaches in \proglang{Stan} which can be easily integrated using the language's \code{\#include} directive. Building on \proglang{Stan} has distinct advantages: First, performant general purpose inference algorithms are implemented and well tested, including ADVI, penalized maximum likelihood estimation, and advanced HMC samplers with automatic hyperparameter tuning. Second, GPs can be used as components in larger hierarchical models without having to adapt or extend the library. Third, \proglang{Stan} supports automatic differentiation obviating the need for implementing gradients manually. Finally, \proglang{Stan} has an engaged community whose members support one another in building statistical models, including extensive expertise in GPs.

In \cref{sec:stan}, we provide a brief introduction to \proglang{Stan} and the \pkg{cmdstanpy} \citep{cmdstanpy} interface for \proglang{Python}. In \cref{sec:graph}, we present an implementation of GPs on directed acyclic graphs which can encode structured dependencies between observations and generalizes nearest-neighbor approximations. In \cref{sec:fourier}, we demonstrate how to use Fourier methods to evaluate the GP likelihood exactly for observations on a regular grid in one and two dimensions. Both implementations are designed to dovetail with \proglang{Stan}'s design philosophy, facilitating their integration into larger models. In \cref{sec:getting-started}, we demonstrate how to use the package with a simple example in both \proglang{Python} and \proglang{R}. We consider a benchmark problem and discuss the importance of different parameterizations for performant inference in \cref{sec:benchmark}. Furthermore, we demonstrate the utility of both approaches with two examples: Inferring the density of trees in a 50~ha plot in Panama \citep{Condit2019} and predicting passenger numbers on the London Underground transportation network. We summarize our contributions in \cref{sec:discussion} and discuss how the package can be employed to build more complex models.

As \proglang{Stan} does not have a package repository, we have published the library as a \proglang{Python} package \pkg{gptools-stan} on PyPI and as an \proglang{R} package \pkg{gptoolsStan} on CRAN. The packages include the \proglang{Stan} library code and provide utility functions to integrate with the popular \proglang{Stan} interfaces \pkg{cmdstanpy} and \pkg{cmdstanr}, respectively. The library can also be obtained directly from \url{https://github.com/onnela-lab/gptools}. Extensive technical documentation and examples are available at \url{https://gptools-stan.readthedocs.org}.

\section[Stan]{Introduction to \proglang{Stan}\label{sec:stan}}

\proglang{Stan} is a probabilistic programming framework, comprising both a concise \proglang{R}-like syntax to declare probabilistic models and an efficient Hamilton Monte Carlo algorithm to draw posterior samples \citep{Betancourt2018}. Readers familiar with \proglang{Stan} may skip to \cref{sec:graph}.

Each \proglang{Stan} program consists of blocks to declare inputs, parameters, and the probabilistic model. For a concrete example, consider a linear regression model with $n\times p$ design matrix $\mat{X}$, coefficient vector $\boldsymbol{\theta}$ with $p$~elements, outcome vector $\vec{y}$ with $n$~elements, and observation noise variance $\sigma^2$, i.e.,
\[
    \vec{y}\dist\mathsf{Normal}\left(\mat{X} \boldsymbol{\theta}, \sigma^2\right).
\]
The corresponding \proglang{Stan} program is shown below.
\begin{CodeInput}
data {
    int n, p;
    matrix [n, p] X;
    vector[n] y;
}

parameters {
    vector[p] theta;
    real<lower=0> sigma;
}

model {
    theta ~ normal(0, 1);
    sigma ~ gamma(2, 2);
    y ~ normal(X * theta, sigma);
}
\end{CodeInput}
The \code{data} block defines inputs required to evaluate the likelihood of the model; \code{parameters} declares parameters of the model including any constraints, such as the noise scale being non-negative. Finally, the \code{model} block declares priors for parameters and the observation model of outcomes $\vec{y}$ given covariates $\mat{X}$ and parameters $\boldsymbol{\theta}$ and $\sigma$. To illustrate the analysis workflow in \proglang{Python}, we generated synthetic data using the \pkg{NumPy} package \citep{Harris2020Numpy} by sampling from the prior predictive distribution with $n=100$ observations and $p=3$ covariates. We fix the random number generator seed for reproducibility.
\begin{CodeInput}
>>> import numpy as np

>>> np.random.seed(0)
>>> n = 100
>>> p = 3
>>> X = np.random.normal(0, 1, (n, p))
>>> theta = np.random.normal(0, 1, p)
>>> sigma = np.random.gamma(2, 2)
>>> y = np.random.normal(X @ theta, sigma)

>>> print(f"coefficients: {theta}")
>>> print(f"observation noise scale: {sigma}")
\end{CodeInput}
\begin{CodeOutput}
coefficients: [-1.307  1.658 -0.118]
observation noise scale: 1.867
\end{CodeOutput}
We used the \pkg{cmdstanpy} interface to compile the above model, draw posterior samples, and report summary statistics.
\begin{CodeInput}
>>> import cmdstanpy

>>> model = cmdstanpy.CmdStanModel(stan_file="linear.stan")
>>> fit = model.sample(data={"n": n, "p": p, "X": X, "y": y}, seed=0)
>>> print(fit.summary())
\end{CodeInput}
\begin{CodeOutput}
               5%      50%      95%   ...
theta[1]   -1.535   -1.229   -0.932   ...
theta[2]    1.439    1.754    2.062   ...
theta[3]   -0.385   -0.062    0.266   ...
sigma       1.715    1.921    2.170   ...
...
\end{CodeOutput}
The 90\% marginal posterior intervals for all parameters are consistent with the values used to generate the data. Having gained some intuition for \proglang{Stan} and \pkg{cmdstanpy}, we consider two approaches to scalable GP inference and their implementation in \proglang{Stan} in the following two sections.

\section[Structured dependencies]{Gaussian processes with structured dependencies\label{sec:graph}}

The joint distribution of observations $\vec{f}$ may be expressed as the product of conditional distributions
\begin{equation}
    \proba{\vec{f}}=\proba{f_1}\prod_{j=2}^n \proba{f_j\mid f_{j-1}, \ldots, f_1}.
    \label{eq:conditional}
\end{equation}
The conditional structure in \cref{eq:conditional} can be encoded by a directed acyclic graph (DAG) whose nodes represent observations such that a directed edge exists from a node $j$ to each of its predecessors $\pred_j=\braces{j-1,\ldots,1}$; the ordering is arbitrary. If two observations do not depend on one another, the corresponding edge can be removed from the DAG to reduce the computational cost. In particular, evaluating each factor of \cref{eq:conditional} requires inverting a matrix with size equal to the number of predecessors of the corresponding node---a substantial saving if the graph is sparse. For example, nearest-neighbor methods, a special case, reduce the asymptotic runtime to $\BigO\parenth{n q^3}$ by retaining only edges from each node to at most $q$ of its nearest predecessors. This approach can yield excellent approximations provided that the neighborhoods are large enough and that the kernel only models local correlations \citep{Wu2022}. For example, nearest-neighbor methods are not suitable for periodic kernels but can be approximated by structured dependencies if the period is known, such as diurnal or yearly patterns.

\begin{algorithm}
\caption{Evaluate the log likelihood of the Gaussian process realization $\vec{f}$ given its mean $\boldsymbol{\mu}$, locations of observations $\vec{x}$, covariance kernel $k$, and the dependency graph encoded as a set of predecessors $\pred$.}\label{alg:graph-lpdf}
\begin{algorithmic}[1]
\Function{gp\_graph\_lpdf}{$\vec{f}\mid \boldsymbol{\mu}, \vec{x}, k, \pred$}
    \State $\mathcal{L} \gets \Call{normal\_lpdf}{f_1\mid\mu_1, k\parenth{x_1,x_1}}$ \Comment{Marginal log likelihood for the first node.}
    \For{$i\in\bracket{2..n}$}\label{alg:graph-lpdf:loop}
    \State $\boldsymbol\Sigma\gets k\parenth{\vec{x}_{\pred_i},\vec{x}_{\pred_i}}$ \Comment{Covariance among predecessors of $i$.}\label{alg:graph-lpdf:Sigma}
    \State $\vec{s}\gets k\parenth{x_i,\vec{x}_{\pred{i}}}$ \Comment{Covariance between $i$ and its predecessors.}
    \State $\nu \gets \vec{s}^\intercal \Sigma^{-1} \boldsymbol\mu_{\pred_i}$ \Comment{Conditional mean.}
    \State $\tau^2=k\parenth{x_i,x_i}- \vec{s}^\intercal \Sigma^{-1}\vec{s}$ \Comment{Conditional variance.}\label{alg:graph-lpdf:tau2}
    \State $\mathcal{L}\gets\mathcal{L} + \Call{normal\_lpdf}{f_i\mid\nu,\tau^2}$ \Comment{Conditional log likelihood for $i$ given $\pred_i$.}
    \EndFor
    \State \Return $\mathcal{L}$
\EndFunction
\end{algorithmic}
\end{algorithm}

Pseudocode to approximate the likelihood of a GP realization $\vec{f}$ using structured dependencies is shown in \cref{alg:graph-lpdf} for a general kernel $k$. The algorithm approximates the log likelihood iteratively by evaluating the conditional mean and variance for each node $i$ given its predecessors $\pred_i$ in lines~\ref{alg:graph-lpdf:Sigma}--\ref{alg:graph-lpdf:tau2}; the conditional distributions are available in closed form for multivariate normal distributions \citep[Appendix~A1]{Gelman2013}. The evaluation of likelihood contributions can be further accelerated by parallelizing the loop in line~\ref{alg:graph-lpdf:loop}.

We implemented a custom distribution in \proglang{Stan} such that a GP with squared exponential kernel on a DAG embedded in a $p$-dimensional space can be specified as
\begin{Code}
f ~ gp_graph_exp_quad_cov(loc, x, sigma, length_scale, edges);
\end{Code}
where \code{vector[n] loc} is the prior mean, and \code{array[n] vector[p] x} is an array of locations in $p$ dimensions for each of the $n$ nodes of the graph. The parameters \code{real sigma} and \code{real length_scale} control the marginal scale and smoothness of the kernel which is defined as \citep[Chapter~2]{Duvenaud2014}
\begin{equation}
    k\parenth{x,x'}=\mathtt{sigma}^2\times\exp\parenth{-\frac{\abs{x-x'}^2}{2\times\mathtt{length\_scale}^2}}.\label{eq:squared-exponential-definition}
\end{equation}
The larger the length scale the more slowly the GP varies because even points with substantial separation $\abs{x-x'}$ remain highly correlated. The graph is encoded by the edge list \code{array[,] int edges}, a two-dimensional array of integer node labels. Each column represents an edge from the node in the second row to the corresponding node in the first row, i.e., edges ``point up'' from successors to predecessors indicating the dependence of the former on the latter. For example, the following edge list represents the directed line graph of four nodes $1\leftarrow2\leftarrow3\leftarrow4$:
\begin{Code}
array [2, 3] int edges = {
    {1, 2, 3},  // Predecessors (where dependency edges end).
    {2, 3, 4}   // Successors (where dependency edges start).
};
\end{Code}
Following \proglang{Stan}'s indexing convention, node labels start at one. Similar distributions are provided for the \code{matern32} and \code{matern52} kernels.


\section[Fourier Gaussian processes]{Gaussian processes in Fourier space\label{sec:fourier}}

We can use Fourier methods to evaluate the likelihood efficiently if three conditions are satisfied \citep[Appendix~B]{Rasmussen2006}. First, we need to consider observation points $\vec{x}$ on a regular grid to reap the computational benefits of the fast Fourier transform (FFT) \citep{Press2007}. Second, the kernel must be stationary, i.e., $k\parenth{x,x'} = k\parenth{x-x'}$ such that the correlation only depends on the separation between observations. Third, the kernel must be $n$-periodic because the FFT is subject to periodic boundary conditions, i.e., $k\parenth{x + n, x'} = k\parenth{x-x'}$, where $n$ is the number of observations. These conditions may seem overly restrictive. However, in many settings, data naturally form a regular grid, e.g., financial time series with fixed sampling interval \citep{Hoffmann2020}, resampled or binned time series \citep{Flaxman2015}, or rasterized images \citep{Tipping2002}. Likewise, stationary kernels, such as squared exponential and Mat\'ern kernels, are common choices for modeling functional data using GPs. Finally, the effect of periodic boundary conditions can be attenuated by padding the domain, as discussed in more detail in \cref{sec:trees,app:padding}.

Because the Fourier transform is a linear operator and $\vec{f}$ is multivariate normal, the discrete Fourier coefficients
\[
    \tilde f_\xi = \sum_{j=0}^{n - 1} \exp\parenth{-\frac{2\pi\imag \xi j}{n}} f_j
\]
are also multivariate normal, where $\xi$ is the (discrete) frequency, $f_j$ is the GP at the $j^\text{th}$ grid point, and $\imag$ is the imaginary number. Assuming $\mu\parenth{x}=0$ for simplicity, the mean of Fourier coefficients is zero and their expected complex-conjugate product at two different frequencies $\xi$ and $\xi'$ is
\begin{align*}
    \E{\tilde f_\xi\overline{\tilde f_{\xi'}}}&=\sum_{j=0}^{n-1}\sum_{j'=0}^{n-1}\exp\parenth{-\frac{2\pi\imag}{n}\parenth{j\xi-j'\xi'}}k\parenth{j-j'}\\
    &=\sum_{j'=0}^{n-1}\exp\parenth{-\frac{2\pi\imag j'}{n}\parenth{\xi-\xi'}}\sum_{\Delta=-j'}^{n-1-j'}\exp\parenth{-\frac{2\pi\imag \Delta}{n}}k\parenth{\Delta},
\end{align*}
where we changed variables to $j=\Delta + j'$ in the second line. The argument of the inner sum is $n$-periodic, and we may shift the limits of summation to $\bracket{0..n-1}$ without changing the sum. The change of limits decouples the two sums. The first is a sum-representation of the Kronecker delta $n\delta_{\xi\xi'}$; the second is the Fourier transform of the kernel $\vec{\tilde k}$. We obtain
\[
    \E{\tilde f_\xi\overline{\tilde f_{\xi'}}}=n\delta_{\xi\xi'}\tilde k_\xi.
\]
Fourier coefficients of different frequencies are thus independent with variance $n \tilde{k}_\xi$.

\begin{algorithm}
\caption{Evaluate the log likelihood of the Gaussian process realization $\vec{f}$ given its mean $\boldsymbol\mu$ and Fourier-transformed covariance kernel $\vec{\tilde k}$. Range indexing is inclusive on the left and exclusive on the right, i.e., $\vec{f}_{a:b}=\braces{f_a,\ldots,f_{b-1}}$.}\label{alg:fourier1d-lpdf}
\begin{algorithmic}[1]
\Function{gp\_rfft\_lpdf}{$\vec{f}\mid \boldsymbol{\mu}, \vec{\tilde k}$}
    \State $\vec{z}\gets \abs{\Call{rfft}{\vec f - \boldsymbol\mu}}$\Comment{Modulus of centered real FFT with $\floor{n / 2} + 1$ elements.}
    \State $\mathcal L\gets \Call{normal\_lpdf}{z_0\mid 0, n\tilde k_0}$ \Comment{Real zero-frequency term.}\label{alg:fourier1d-lpdf:zero-freq}
    \If{$n \bmod 2 = 1$}
        \State $m\gets \frac{n + 1}{2}$\Comment{Index following highest-frequency complex coefficient.}
    \Else
        \State $m\gets \frac{n}{2}$\Comment{Index following highest-frequency complex coefficient.}
        \State $\mathcal L\gets \Call{normal\_lpdf}{ z_{m}\mid 0, n\tilde k_{m}}$\Comment{Real Nyquist-frequency term.}\label{alg:fourier1d-lpdf:nyquist-freq}
    \EndIf
    \State $\mathcal L\gets \mathcal L + 2\times \Call{normal\_lpdf}{\vec{z}_{1:m}\mid 0, n\vec{\tilde k}_{1:m}}$\Comment{Complex oscillatory terms.}\label{alg:fourier1d-lpdf:oscillatory}
    \State \Return $\mathcal{L}$
\EndFunction
\end{algorithmic}
\end{algorithm}

Subject to careful bookkeeping, we can evaluate the likelihood exactly, as illustrated in \cref{alg:fourier1d-lpdf}. Because $\vec f$ is real, we use the real FFT (RFFT) for efficiency. It comprises $\floor{n/2}+1$ complex coefficients because just under half the coefficients are redundant \citep[Chapter~12.3]{Press2007}. The zero-frequency term $\tilde f_0$ and, for even $n$, the Nyquist frequency term $\tilde f_{n / 2}$ are real (see lines~\ref{alg:fourier1d-lpdf:zero-freq} and~\ref{alg:fourier1d-lpdf:nyquist-freq} of \cref{alg:fourier1d-lpdf}, respectively). The complex coefficients contribute twice in line~\ref{alg:fourier1d-lpdf:oscillatory} to account for the redundant terms omitted by the RFFT.

We implemented a custom distribution in \proglang{Stan} such that a GP on a grid can be specified as
\begin{Code}
f ~ gp_rfft(loc, cov_rfft);
\end{Code}
where \code{vector[n] loc} is the prior mean and \code{vector[n \%/\% 2 + 1] cov_rfft} is the RFFT of the kernel evaluated on the grid (\code{\%/\%} denotes floor division in \proglang{Stan}).

\begin{figure}
    \centering
    \includegraphics{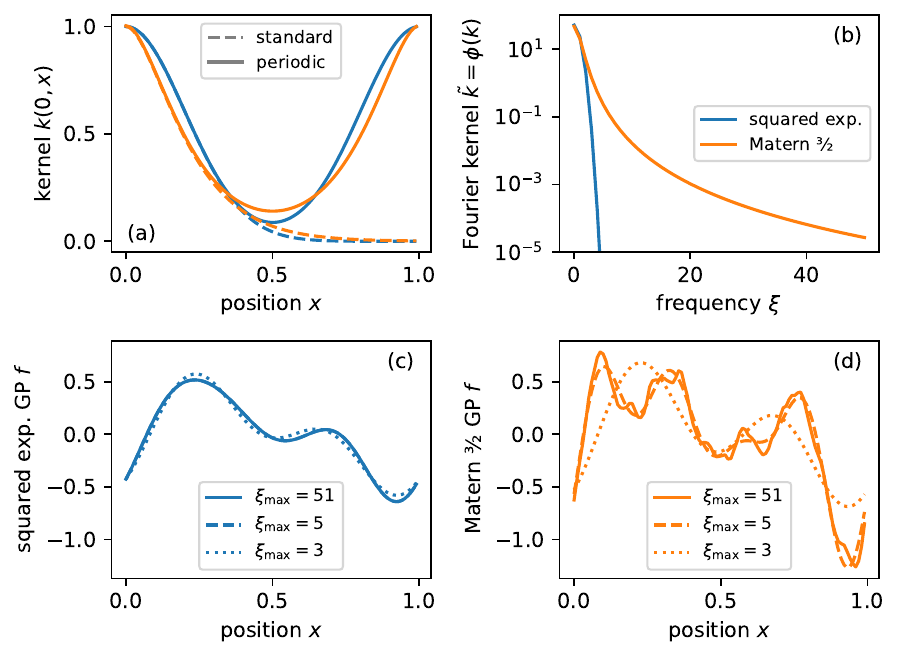}
    \caption{Gaussian processes can often be captured by a small number of Fourier modes. Panel~(a) shows the periodic and non-periodic (standard) versions of the squared exponential kernel (blue) and Mat\'ern~$\sfrac{3}{2}$ kernel (orange) with $\sigma=1$ and $\ell=0.2$. The two versions are hardly distinguishable for $x<\ell$. Padding may need to be introduced if non-periodic signals are modeled with periodic kernels. The power spectrum of the two periodic kernels is shown in panel~(b). The squared exponential kernel has negligible power for all but the first few frequencies, explaining why it is often considered too smooth to ``represent natural phenomena'' \citep{Handcock1993}. Panels~(c) and~(d) show realizations of GPs with squared exponential and Mat\'ern~$\sfrac{3}{2}$ kernels, respectively. Different line styles correspond to approximations with a different number of Fourier modes. For the squared exponential kernel, the number of modes can be reduced by an order of magnitude without substantially affecting realizations. The Mat\'ern kernel requires more modes due to its heavy-tailed power spectrum.}
    \label{fig:kernels}
\end{figure}

Fortunately, the RFFT of common kernels, such as the squared exponential kernel and Mat\'ern kernels, can be evaluated directly in the Fourier domain \citep[Chapter~4]{Rasmussen2006}, as shown in \cref{fig:kernels} (see \cref{app:kernel-expressions} for definitions of the kernel). Evaluating the kernel in the Fourier domain also obviates the need for small ``nugget'' variance or ``jitter'' typically required for numerical stability \citep{Neal1997}. We thus only need to evaluate one Fourier transform, that of the signal, to evaluate the likelihood. The library provides the following functions to evaluate Fourier-domain kernels:
\begin{Code}
gp_periodic_exp_quad_cov_rfft(n, sigma, length_scale, period)
gp_periodic_matern_cov_rfft(n, nu, sigma, length_scale, period)
\end{Code}
where \code{period} is the size of the domain, \code{n} is the number of grid points with spacing \code{period / n}, and \code{nu} is the smoothness parameter of the Mat\'ern kernel. For \code{nu = 0.5}, the Mat\'ern kernel reduces to the exponential kernel $k\parenth{x,x'}=\sigma^2\exp\parenth{\frac{\abs{x-x'}}{\mathtt{length\_scale}}}$, and, in the limit of large \code{nu}, it converges to the squared exponential kernel defined in \cref{eq:squared-exponential-definition} \citep[Chapter~4.2]{Rasmussen2006}. The parameters \code{sigma} and \code{length_scale} have the same meaning as in \cref{sec:graph}. Equivalent functions, which we discuss further in \cref{sec:trees}, are provided for two-dimensional grids. We implemented the Fourier-domain kernels by naively discretizing frequencies. This approach works well if the number of grid points is large and the correlation length is small compared with the size of the domain. More sophisticated methods may be required otherwise \citep{Borovitskiy2020}.

\section[Getting started]{Getting started}\label{sec:getting-started}

We demonstrate how to use \pkg{gptools} using a simple example: Drawing samples from a Gaussian process prior using the Fourier methods discussed in \cref{sec:fourier}. The \proglang{Stan} model comprises five blocks:
\begin{Code}
functions {
    #include gptools/util.stan
    #include gptools/fft.stan
}

data {
    int n;
    real<lower=0> sigma, length_scale, period;
}

transformed data {
    vector [n 
        gp_periodic_exp_quad_cov_rfft(n, sigma, length_scale, period) + 1e-9;
}

parameters {
    vector [n] f;
}

model {
    f ~ gp_rfft(zeros_vector(n), cov_rfft);
}
\end{Code}
The first block \code{functions} includes the source of \pkg{gptools} in the program, and the second block \code{data} declares the number of grid points \code{n} and kernel parameters (see \cref{sec:fourier} for definitions). Because the kernel parameters are fixed, we can precompute the RFFT of the kernel in the \code{transformed data} block. Finally, we declare the vector \code{f} representing the Gaussian process in the \code{parameters} block and specify the prior in the \code{model} block.

\subsection[Getting started in Python]{Getting started in \proglang{Python} using \pkg{cmdstanpy}}

To install \pkg{gptools} for \proglang{Python}, run
\begin{Code}
$ pip install gptools-stan
\end{Code}
from the command line. This will download the \proglang{Stan} source and install a \proglang{Python} package with utility functions to compile models using the \pkg{cmdstanpy} interface \citep{cmdstanpy}. Executing the following \proglang{Python} script will compile and fit the model:
\begin{Code}
>>> import cmdstanpy
>>> from gptools.stan import get_include
>>>
>>> model = cmdstanpy.CmdStanModel(
...     stan_file="getting_started.stan",
...     stanc_options={"include-paths": get_include()},
... )
>>> fit = model.sample(
...     data = {"n": 100, "sigma": 1, "length_scale": 0.1, "period": 1},
...     chains=1,
...     iter_warmup=500,
...     iter_sampling=50,
... )
>>> fit.f.shape
(50, 100)
\end{Code}
The function \code{get_include} returns the path to the \proglang{Stan} source files of \pkg{gptools}. The path is passed to \code{cmdstanpy.CmdStanModel}, which compiles the model, as \code{stanc_options}.
The object \code{fit.f} is a two-dimensional array comprising 50 samples of the vector \code{f} each having 100 elements.
If this is the first time \pkg{cmdstanpy} is used, \pkg{cmdstan} may need to be installed by running
\begin{Code}
$ python -m cmdstanpy.install_cmdstan
\end{Code}
from the command line to install \pkg{cmdstan} \citep{cmdstan} before the model can be compiled and fit.

\subsection[Getting started in R]{Getting started in \proglang{R} using \pkg{cmdstanr}}

To install \pkg{gptools} for \proglang{R}, run
\begin{Code}
> install.packages(
+   "cmdstanr",
+   repos = c("https://mc-stan.org/r-packages/", getOption("repos"))
+ )
> install.packages("gptoolsStan")
\end{Code}
from the \proglang{R} console. The first command installs the \pkg{cmdstanr} interface \citep{cmdstanr} which is not yet available on CRAN, and the second installs \pkg{gptools} for \proglang{R}. Executing the following \proglang{R} script will compile and fit the model:
> \begin{Code}
> library(cmdstanr)
> library(gptoolsStan)
>
> model <- cmdstan_model(
+   stan_file="getting_started.stan",
+   include_paths=gptools_include_path(),
+ )
> fit <- model$sample(
+   data=list(n=100, sigma=1, length_scale=0.1, period=1),
+   chains=1,
+   iter_warmup=500,
+   iter_sampling=50
+ )
> f <- fit$draws("f")
> dim(f)
[1]  50   1 100
\end{Code}
The function \code{gptools_include_path} returns the path to the \proglang{Stan} source files of \pkg{gptools}. The path is passed to \code{cmdstan_model} which compiles the model.
The object \code{f} is an array of samples from the prior distribution. If this is the first time \pkg{cmdstanr} is used, \pkg{cmdstan} may need to be installed by running
\begin{Code}
> install_cmdstan()
\end{Code}
from the \proglang{R} console to install \pkg{cmdstan} \citep{cmdstan} before the model can be compiled and fit.

\section[Benchmark and parameterizations]{Benchmark and the importance of parameterizations\label{sec:benchmark}}

We consider a simple benchmark problem to study the performance of different methods and compare them with standard Gaussian process inference which inverts the kernel. The model comprises a one-dimensional zero-mean Gaussian process prior with squared exponential kernel and an independent normal observation model with variance $\kappa^2$, i.e.,
\begin{equation}\begin{aligned}
    \vec{f}&\dist\mathsf{MultivariateNormal}\parenth{0, \mat{K}}\\
    \vec{y}&\dist\mathsf{Normal}\parenth{\vec{f}, \kappa^2}.
\end{aligned}\label{eq:benchmark-centered}\end{equation}
We used a marginal kernel scale $\sigma=1$ and unit correlation length $\ell=1$ to evaluate the covariance matrix $\mat{K}$ on an integer grid, i.e., $\vec{x}=\braces{0, \ldots, n - 1}$.
Employing the \pkg{cmdstanpy} interface, we drew 100~posterior samples each from 20 independent chains after 100~warmup samples. Warmup samples are used to adapt the sampler for efficient exploration of the posterior \citep{Homan2014}. Default values were used for all other parameters. We considered different dataset sizes between $n=2^4$ and $n=2^{14}$ and allocated a maximum computational budget of one minute for each chain and all $n$, i.e., individual chains were terminated if they did not complete after 60~seconds.

\begin{figure}
    \centering
    \includegraphics{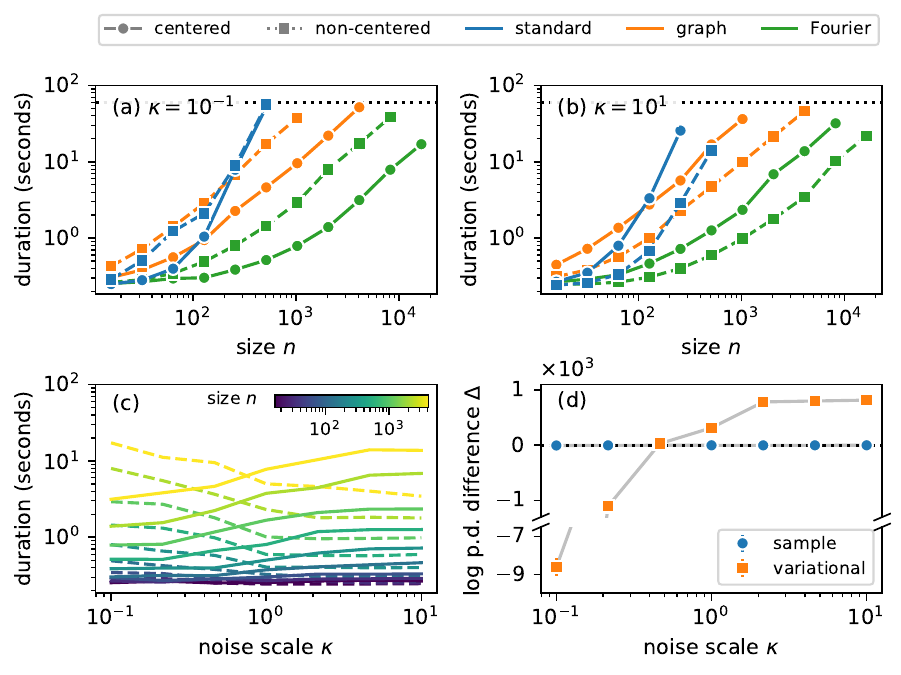}
    \caption{Different approaches, parameterizations, as well as the informativeness of the data substantially affect runtimes. Centered parameterizations are preferable when the data are strong (small observation noise scale $\kappa$), and non-centered parameterizations are superior when the data are weak (large $\kappa$), as shown in panels~(a) and~(b), respectively. Independent of parameterization and dataset size, Fourier methods offer the best performance when observations are regularly spaced. For datasets exceeding a few hundred observations, graph methods are faster than the standard approach which requires inversion of the covariance matrix. The dotted horizontal line represents the maximum computational budget of 60~s. As shown in panel~(c) for the Fourier approach, the runtime of the centered and non-centered parameterizations increases and decreases, respectively, as the data become less informative. While different parameterizations are primarily a performance concern for drawing posterior samples, they have important consequences for variational posterior approximations. Panel~(d) shows the difference $\Delta$ in log posterior density (p.d.) between the non-centered and centered parameterization on 20\% held-out data for $n=1{,}024$ using the Fourier approach. If the model is fit with a variational mean-field approximation, the non-centered parameterization offers higher log posterior scores than the centered parameterization when the data are weak and vice versa. Bootstrapped standard errors are smaller than the size of markers in panel~(d).}
    \label{fig:scaling}
\end{figure}

The mean runtime as a function of dataset size is shown in panels~(a) and~(b) of \cref{fig:scaling} for small ($\kappa=0.1$) and large ($\kappa=10$) noise scales as solid lines, respectively. As expected, the runtime of the standard approach grows rapidly as $n$ increases. We observed an empirical runtime scaling of $n^{\approx 2.5}$ for the standard approach, not dissimilar from the expected asymptotic scaling of $\BigO\parenth{n^3}$. Exploring models with more than a few hundred data points is prohibitively expensive---even for this simple setup. For the graph-based approach, we used the five nearest predecessors ($q=5$) to construct a dependency graph. The method is comparatively slow for small datasets but outperforms the standard approach as $n$ grows. The Fourier approach has the best performance irrespective of dataset size but is limited to observations on a grid.

The model in \cref{eq:benchmark-centered} employs the natural \emph{centered} parameterization \citep{Papaspiliopoulos2007}, i.e., each observation $y_i$ is independent given the corresponding latent $f_i$. This parameterization works well if the data are informative (small $\kappa$) because each observation $y_i$ constrains the corresponding latent parameter $f_i$. The elements of $\vec{f}$ are thus relatively uncorrelated under the posterior, and the Hamiltonian sampler can explore the distribution efficiently~\citep{Homan2014}.

However, if the data are weak (large $\kappa$), they cannot independently constrain each element of $\vec{f}$ and the GP prior dominates the posterior. The resulting correlation among elements of $\vec{f}$ frustrates the sampler, especially if the correlation length is large. We can overcome this challenge by employing a \emph{non-centered} parameterization such that the parameters of the model are uncorrelated under the prior \citep{Papaspiliopoulos2007}. Here, we reparameterize the model in terms of a white noise vector $\vec{z}$ of the same size as $\vec{f}$ and obtain realizations of the GP $\vec{f}=\phi^{-1}\parenth{\vec{z},\boldsymbol{\mu},\mat{K}}$ using an inverse transform $\phi^{-1}$ which must be selected carefully to ensure $\vec{f}$ follows the desired distribution. We chose the inverse transform for consistency with the FFT: The forward transform maps to the Fourier domain, and the inverse transform maps to real space. The reparameterized model is
\begin{equation}\begin{aligned}
    \vec{z}&\dist\mathsf{Normal}\parenth{0, 1}\\
    \vec{f}&=\phi^{-1}\parenth{\vec{z}, 0, \mat{K}}\\
    \vec{y}&\dist\mathsf{Normal}\parenth{\vec{f}, \kappa^2}.
\end{aligned}\label{eq:benchmark-non-centered}\end{equation}

We implemented the following transforms for the graph-based and Fourier approaches:
\begin{Code}
f = gp_inv_graph_exp_quad_cov(z, loc, x, sigma, length_scale, edges);
f = gp_inv_rfft(z, loc, cov_rfft);
\end{Code}
where \code{vector[n] z} are the non-centered white noise parameters and all other parameters are as described previously. For the standard method, we implemented the non-centered parameterization as $\vec{f}=\mat{L}\vec{z}$ \citep{Papaspiliopoulos2007}, where $\mat{L}$ is the Cholesky decomposition of the covariance matrix~$\mat{K}$ such that $\mat{K}=\mat{L}\mat{L}^\intercal$.

\begin{algorithm}
    \caption{Transform white noise $\vec{z}$ to a Gaussian process realization $\vec{f}$ given its mean $\boldsymbol{\mu}$, locations of observations $\vec{x}$, covariance kernel $k$, and the dependency graph encoded as a set of predecessors $\pred$.}
    \label{alg:gp_inv_graph}
    \begin{algorithmic}[1]
        \Function{gp\_inv\_graph}{$\vec{z},\boldsymbol{\mu},\vec{x},k,\pred$}
        \State $f_1\gets \mu_1 + \sqrt{k\parenth{x_1,x_1}}z_1$ \Comment{Sample first observation from the marginal distribution.}\label{alg:gp_inv_graph:marginal}
        \For{$i\in\bracket{2..n}$}\label{alg:gp_inv_graph:loop}
            \State \ldots \Comment{Compute conditional mean $\nu$ and variance $\tau^2$ as in \cref{alg:graph-lpdf}.}
            \State $f_i\gets \nu + \tau z_i$ \Comment{Sample from the conditional distribution given $\pred_i$.}\label{alg:gp_inv_graph:conditional}
        \EndFor
        \State\Return $\vec{f}$
        \EndFunction
    \end{algorithmic}
\end{algorithm}

\Cref{alg:gp_inv_graph} implements the transform for approximate GPs using structured dependencies and closely follows \cref{alg:graph-lpdf}. Instead of evaluating the log likelihood iteratively given the conditional distribution, the algorithm draws a sample sequentially by transforming white noise to the target distribution given previous samples in lines~\ref{alg:gp_inv_graph:marginal} and~\ref{alg:gp_inv_graph:conditional}. Unlike \cref{alg:graph-lpdf} the loop in line~\ref{alg:gp_inv_graph:loop} cannot be parallelized because the conditional mean and variance depend on the results of previous iterations.

\begin{algorithm}
    \caption{Transform white noise $\vec z$ to a Gaussian process realization $\vec{f}$ given its mean $\boldsymbol\mu$ and Fourier-transformed covariance kernel $\vec{\tilde k}$. Range indexing is inclusive on the left and exclusive on the right, i.e., $\vec{f}_{a:b}=\braces{f_a,\ldots,f_{b-1}}$.}
    \label{alg:gp_inv_rfft}
    \begin{algorithmic}[1]
        \Function{gp\_inv\_rfft}{$\vec{z},\boldsymbol{\mu}, \vec{\tilde k}$}
            \State $\boldsymbol{\tau} = \sqrt{n \vec{\tilde k}}$ \Comment{Evaluate standard deviation of Fourier coefficients.}
            \State $\tilde f_0\gets \tau_0 z_0$ \Comment{Real zero-frequency term.}\label{alg:gp_inv_rfft:zero}
            \If{$n \bmod 2 = 1$}
                \State $m\gets\frac{n+1}{2}$ \Comment{Index following highest-frequency complex coefficient.}
            \Else
                \State $m\gets\frac{n}{2}$ \Comment{Index following highest-frequency complex coefficient.}
                \State $\tilde f_m\gets \tau_m z_m$ \Comment{Real Nyquist-frequency term.}\label{alg:gp_inv_rfft:nyquist}
            \EndIf
            \State$\vec{\tilde f}_{1:m}\gets \boldsymbol{\tau}_{1:m}\parenth{\vec{z}_{1:m}+\imag\times\vec{z}_{m + \parenth{n + 1}\bmod 2:n}}/\sqrt2$\Comment{Complex oscillatory terms.}\label{alg:gp_inv_rfft:complex}
            \State$\vec{f}\gets\Call{inv\_rfft}{\vec{\tilde f},n}$ \Comment{Inverse RFFT returning a vector with $n$ elements.}
            \State\Return $\vec{f}$
        \EndFunction
    \end{algorithmic}
\end{algorithm}

The transformation from white noise to a GP realization using Fourier methods is illustrated in \cref{alg:gp_inv_rfft}. As in \cref{alg:fourier1d-lpdf}, we account for the real zero-frequency term and, for even $n$, Nyquist frequency term in lines~\ref{alg:gp_inv_rfft:zero} and~\ref{alg:gp_inv_rfft:nyquist}, respectively. Line~\ref{alg:gp_inv_rfft:complex} constructs the $\floor{\parenth{n-1}/2}$ complex Fourier coefficients from $\vec{z}_{1:m}$ (real part) and $\vec{z}_{m + \parenth{n + 1}\bmod 2:n}$ (imaginary part). The $\parenth{n + 1}\bmod 2$ term in the index accounts for the presence of the Nyquist frequency for even $n$. The omission of redundant terms in the RFFT is addressed by dividing the complex coefficients by $\sqrt 2$.

As shown in panels~(a) and~(b) of \cref{fig:scaling}, the non-centered parameterization (dashed lines) is more performant than the centered parameterization (solid lines) if the noise scale is large and vice versa. Panel~(c) further illustrates the importance of choosing the right parameterization: The runtime differs by up to a factor of five as we vary the noise scale $\kappa$. The higher-frequency terms of smooth GPs have low power, as shown in panel~(b) of \cref{fig:kernels}. We can further improve performance of the non-centered parameterization by discarding all but the first few low-frequency terms. This approach is particularly effective for the squared exponential kernel because the power spectrum decays rapidly with increasing frequency. For the example shown in panel~(c), a GP using only the first five Fourier modes is indistinguishable from the GP considering all 51~modes, reducing the dimensionality of the parameter space by an order of magnitude. GPs with Mat\'ern kernels typically require more Fourier modes because the power spectrum has a relatively heavy tail.

While parameterization is primarily a performance concern for Hamiltonian Monte Carlo samplers, it can have a substantial impact on the predictive ability of models if variational mean-field inference is used. Variational approximations of the posterior tend to assign low probability mass to regions of the parameter space where the full posterior has low mass \citep[Chapter~10.1]{Bishop2006}. Consequently, variational approximations are too narrow if the posterior is highly correlated, and we expect predictions to be overconfident. To test this hypothesis, we sampled synthetic data with $n=1{,}024$ data points from the prior predictive distribution and fitted the models in \cref{eq:benchmark-centered,eq:benchmark-non-centered} to 80\% of each synthetic dataset using variational inference \citep{Kucukelbir2017}. \proglang{Stan}'s ADVI implementation approximates the posterior by a product of independent normal distributions, one for each parameter. The variational approximation is optimized in an unconstrained space. Constrained parameters are obtained by applying a transform, e.g., an exponential transform to obtain a positive parameter such as the length scale $\ell$. We repeated the analysis 20 times for each noise scale $\kappa$, terminating the algorithm, as before, if it did not complete after one minute.

We evaluated the predictive ability of the fitted models by evaluating the log posterior density on the 20\% held-out GP realizations, i.e., $\log\proba{\vec{f}_\text{test}\mid \vec{y}_\text{train}}$. To compare the parameterizations, we approximated the log posterior difference $\Delta=\log\proba{\vec{f}_\text{test}=\phi^{-1}\parenth{\vec{z}_\text{test}}\mid \vec{y}_\text{train}}-\log\proba{\vec{f}_\text{test}\mid \vec{y}_\text{train}}$ using a Gaussian kernel density estimator \citep[Chapter~2.5.1]{Bishop2006}.
The log posterior differences $\Delta$ were averaged over the 20 independent fits (or fewer if the algorithm failed to complete within the allocated computational budget).
As shown in panel~(d) of \cref{fig:scaling}, the non-centered parameterization makes better predictions than the centered parameterization when the noise scale is large and vice versa. The error bars shown represent bootstrapped standard errors, i.e., the standard deviation of $\Delta$ resampled from the population of independent chains \citep{Rubin1981Bootstrap}.
We also fitted the two parameterizations by drawing posterior samples using \proglang{Stan}'s Hamiltonian sampler and evaluated the log posterior difference. Different parameterizations did not affect the predictive performance but had a significant impact on runtime.

\section[Illustrations]{Illustrations\label{sec:illustration}}

\subsection[London Underground]{Passengers on the London Underground transportation network\label{sec:tube}}

\begin{figure}
    \centering
    \includegraphics{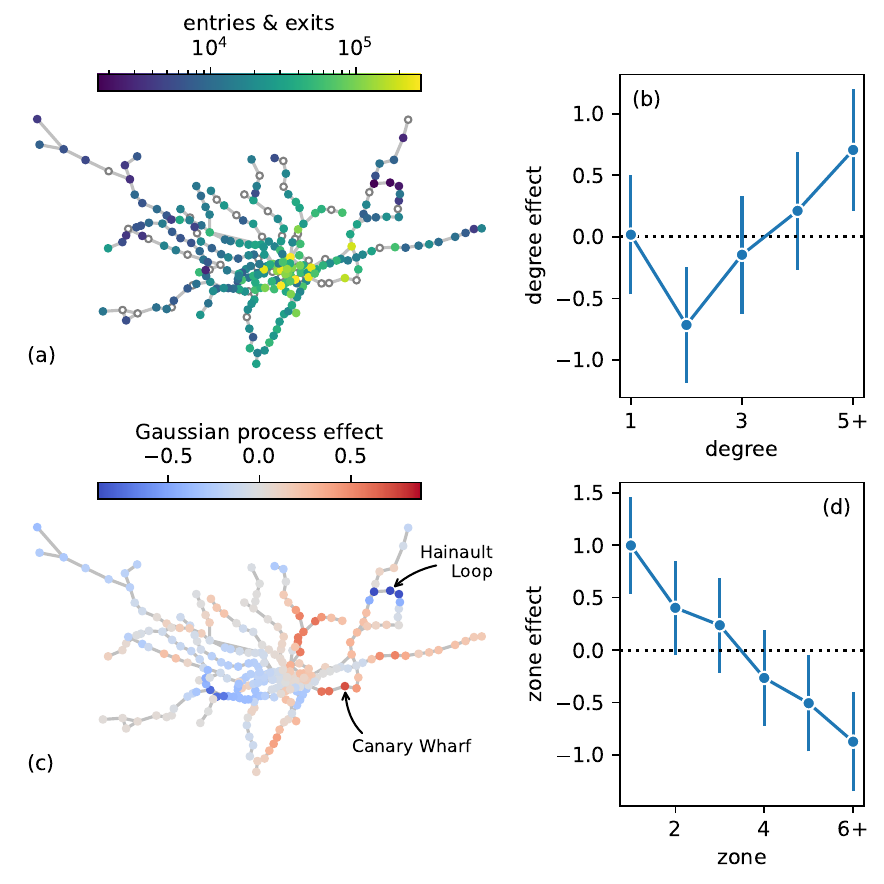}
    \caption{A Gaussian process on the London Underground transportation network identifies idiosyncrasies of transport use. Panel~(a) shows the daily average number of entries and exits for each of the 267~stations in 2019 together with the transport network. Empty nodes denote held-out data. Panels~(b) and~(d) show the effect of degree and zone on passenger numbers, respectively. Central stations with larger degree tend to have more passengers. Termini with degree one have uncharacteristically many passengers because they serve as interchanges for longer-distance trains. Panel~(c) shows the GP which explains residual variations in passenger volumes after controlling for zone and degree. On the one hand, stations in the Hainault loop have relatively few passengers because they are served infrequently compared with nearby stations. On the other hand, Canary Wharf has surprisingly many entries and exits because it is a busy financial center in London.}
    \label{fig:tube}
\end{figure}

Millions of people use the London Underground transportation network, commonly referred to as the ``Tube'', to travel across the city each day \citep{Crowding}. The number of passengers using each station is affected by various factors, including how connected it is and which zone the station is in (the network comprises nine transport zones). In addition to these fixed effects, we also expect passenger numbers to be affected by smooth spatial effects, e.g., due to variability in population density. The spatial effect can naturally be modeled as a GP with structured dependencies induced by the transport network itself. We converted the undirected graph to a directed acyclic graph in two steps. First, we assigned an integer label to each node. Second, we added an edge from node $i$ to node $j$ if $i < j$ and the corresponding edge exists in the undirected graph. The order of nodes is arbitrary because the joint probability in \cref{eq:conditional} can be factorized in any order. We collected network data from the Transport for London open data API \citep{TfLAPI} and obtained the average daily number of entries and exits at each station in 2019 \citep{Crowding}, as shown in panel~(a) of \cref{fig:tube}. The model includes fixed effects for each zone and degree, i.e., the number of connections a station has. The corresponding regression coefficients were mildly regularized by half-t priors with two degrees of freedom \citep{Gelman2008}. We truncated the degree and zone of each station at five and six, respectively, because only few stations exceed these values. The overall number of passengers is captured by a scalar $\mu$, and we used the GP to explain any residual effects. A squared exponential kernel was employed for the covariance, and we used a half-t prior for the marginal scale. The correlation length of the kernel is not identifiable if it is smaller than the smallest distance between stations (0.16~km) or larger than the extent of the transportation network (62~km) \citep{Trangucci2016}. We thus used a log-uniform prior on the interval $\bracket{0.32\,\text{km},31\,\text{km}}$ to suppress extreme length scales. All distances were evaluated in the Ordnance Survey National Grid projection (\code{epsg:27700}). A non-centered parameterization was used because the residual effects are not strongly identified by the data after controlling for zone and degree. We used a log-normal observation model (rather than a model for count data) because passenger data are heavy-tailed and reported as daily averages.
\begin{CodeInput}
functions {
    #include gptools/util.stan
    #include gptools/graph.stan
}

data {
    int num_stations, num_edges, num_zones, num_degrees;
    array[num_stations] vector[2] station_locations;
    array[num_stations] int passengers;
    array[2, num_edges] int edge_index;
    matrix[num_stations, num_zones] one_hot_zones;
    matrix[num_stations, num_degrees] one_hot_degrees;
}

parameters {
    vector[num_stations] z;
    real mu;
    real<lower=0> sigma, kappa;
    real<lower=log(0.32), upper=log(31)> log_length_scale;
    vector[num_zones] zone_effect;
    vector[num_degrees] degree_effect;
}

transformed parameters {
    real length_scale = exp(log_length_scale);
    vector[num_stations] f = gp_inv_graph_exp_quad_cov(
        z, zeros_vector(num_stations), station_locations, sigma,
        length_scale, edge_index);
    vector[num_stations] log_mean = mu + f + one_hot_zones
        * zone_effect + one_hot_degrees * degree_effect;
}

model {
    z ~ std_normal();
    sigma ~ student_t(2, 0, 1);
    zone_effect ~ student_t(2, 0, 1);
    degree_effect ~ student_t(2, 0, 1);
    kappa ~ student_t(2, 0, 1);
    for (i in 1:num_stations) {
        if (passengers[i] > 0) {
            log(passengers[i]) ~ normal(log_mean[i], kappa);
        }
    }
    // We use an implicit uniform prior on `log_length_scale`.
}
\end{CodeInput}

We fitted the model to 80\% of the passenger data using the below \proglang{Python} code, withholding 20\% of the stations uniformly at random for later evaluation. Held-out data are encoded as \code{-1} in the \proglang{Stan} model.
\begin{CodeInput}
>>> from gptools.stan import get_include
>>> import json
>>> import numpy as np

>>> # Load station locations, edges, passenger numbers, apply training mask.
>>> with open("tube-stan.json") as fp:
...     data = json.load(fp)
>>> train_mask = np.random.binomial(1, 0.8, data["num_stations"])
>>> data["passengers"] = np.where(train_mask, data["passengers"], -1)

>>> # Compile model and fit it.
>>> model = compile_model(
...     stan_file="tube.stan",
...     stanc_options={"include-paths": get_include()},
... )
>>> fit = model.sample(data)
>>> print(fit.diagnose())
\end{CodeInput}
\begin{CodeOutput}
Processing csv files: ...

Checking sampler transitions treedepth.
Treedepth satisfactory for all transitions.

Checking sampler transitions for divergences.
No divergent transitions found.

Checking E-BFMI - sampler transitions HMC potential energy.
E-BFMI satisfactory.

Effective sample size satisfactory.

Split R-hat values satisfactory all parameters.

Processing complete, no problems detected.
\end{CodeOutput}

We employed the default configuration of \pkg{cmdstanpy} to draw posterior samples, resulting in four independent chains with 2,000~samples each. The \code{print(fig.diagnose())} call evaluates and reports a suite of diagnostics to identify potential problems and assess convergence. For example, the split $\hat R$ statistic compares samples both within and between chains to determine whether they are likely to have mixed well \citep{Vehtari2021}. Tree depth, divergence, and Bayesian fraction of missing information (BFMI) are technical diagnostics to assess whether the sampler was able to explore the posterior distribution; the effective sample size estimates the number of independent samples drawn which may be smaller than 2,000 due to autocorrelation within each chain \citep{Betancourt2018}. Here, the samples satisfied all posterior checks offered by \pkg{cmdstan} \citep{CmdStanUserGuide}.

Panel~(b) shows the effect of degree on passenger numbers on the log scale. They tend to increase with the degree of a station as they offer passengers a variety of travel options. Termini with degree one are an exception: Their passenger numbers are uncharacteristically large because they serve as stepping stones to longer-distance travel beyond the Tube network. Unsurprisingly, central stations in zones one to three tend to have more passengers than stations in the suburbs (zones four and above), as shown in panel~(d). The GP captures any residuals that cannot be explained by the degree of the station or the zone it is located in, as shown in panel~(c). For example, on the one hand, Canary Wharf has the largest residual effect. It is one of London's financial centers, and the station serves tens of thousands of commuters each day despite being a station without an interchange. On the other hand, stations in the north of the Hainault loop have the largest negative residual effect because the stations are served by only three trains an hour \citep{CentralLineTimetable}. Passengers divert to nearby stations that are served by twelve trains an hour \citep{CentralLineTimetable}. Comparing the model with a model without GP effects using the log posterior predictive distribution on held-out data, we observe no significant difference after bootstrapping errors. Nevertheless, this example illustrates how our package can be used to easily construct GPs with structured dependencies. It may be necessary to obtain an edge list from geospatial data, e.g., for the analysis of spatially correlated outcomes \citep{Morris2019BYM}. The best approach depends on the problem at hand and how many neighbors are considered for each geographical unit.  Common packages for manipulating geospatial data include \pkg{shapely} \citep{ShapelyPackage} and \pkg{geopandas} \citep{GeopandasPackage} for \proglang{Python} and \pkg{spdep} \citep{Bivand2022GeospatialRPackages} for \proglang{R}.

\subsection[Tree density]{Density of \emph{T. panamensis} on a 50~ha plot in Panama\label{sec:trees}}

\begin{figure}
    \centering
    \includegraphics{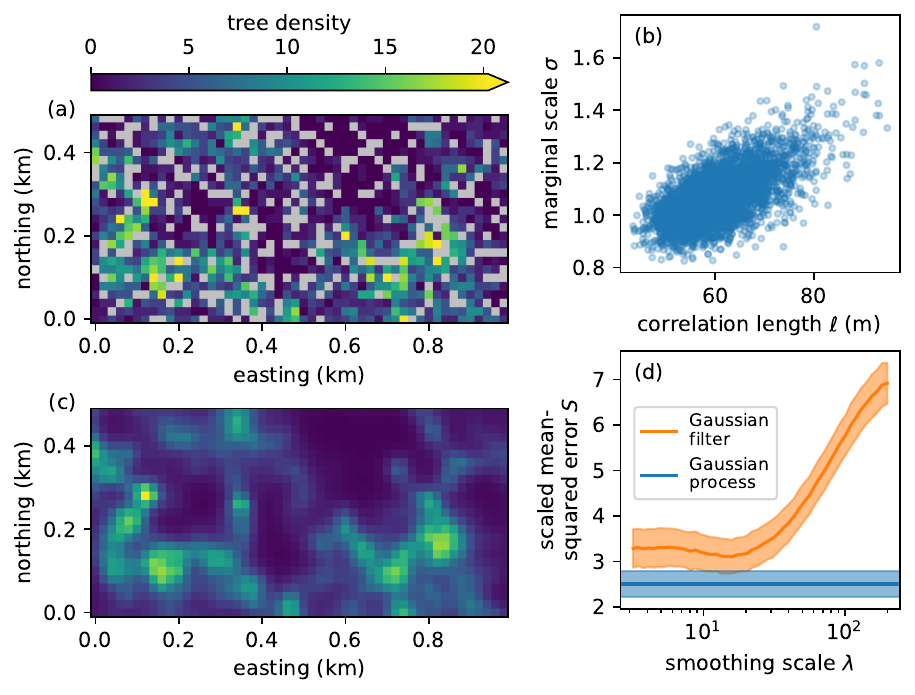}
    \caption{A Gaussian process based on the two-dimensional Fourier transform can accurately predict the frequency of trees. Panel~(a) shows the number of \emph{T. panamensis} trees per 20~m quadrant collected during the 2015~census of the 50~ha Barro Colorado plot in Panama~\citep{Condit2019} as a heat map. Gray quadrants indicate the 20\% held-out test data. Panel~(c) shows the posterior median of the expected tree frequency, recovering the held-out data and smoothing the empirical frequencies. Posterior samples of the correlation length and marginal scale of the Mat\'ern kernel are shown in panel~(b). Correlation length samples are well below the padding (200~m) introduced to mitigate the effect of periodic boundary conditions. Panel~(d) shows the scaled mean squared error with bootstrapped errors on the held-out data under the Gaussian process model and a Gaussian filter with variable smoothing scale.}
    \label{fig:trees}
\end{figure}

To illustrate the use of Fourier methods, we consider the density of \emph{T. panamensis} trees during the 2015~census of the 50~ha Barro Colorado plot in Panama~\citep{Condit2019}. The plot is divided into quadrants of 20~m side length. As shown in panel~(a) of \cref{fig:trees}, the data comprise the frequency of trees within each quadrant, i.e., a matrix of count data with shape $\parenth{25, 50}$. The observed tree frequency counts $\vec{y}$ were modeled by a negative-binomial distribution to account for possible overdispersion. We used a latent GP to model the log-mean of this distribution and capture the tree density. We employed a Mat\'ern kernel with smoothness parameter $\nu=\sfrac{3}{2}$, half-t prior for the marginal scale and overdispersion parameter, and log-uniform prior for the correlation length as in \cref{sec:tube}. Because the quadrants are regularly spaced, the likelihood can be evaluated exactly using Fourier methods, as discussed in \cref{sec:fourier}. However, unlike the FFT, trees are not subject to periodic boundary conditions. To mitigate this issue and reduce correlation between opposing sides of the plot, we padded the matrix with ten additional quadrants in each dimension (corresponding to 200~m) resulting in a matrix with shape $\parenth{35, 60}$. Despite increasing the number of latent variables by almost 70\%, the method is faster than the standard approach which inverts the covariance matrix.
\begin{Code}
functions {
    #include gptools/util.stan
    #include gptools/fft.stan
}

data {
    int num_rows, num_cols, num_rows_padded, num_cols_padded;
    array[num_rows, num_cols] int frequency;
}

parameters {
    matrix[num_rows_padded, num_cols_padded] z;
    real mu;
    real<lower=0> sigma, kappa;
    real<lower=log(2), upper=log(28)> log_length_scale;
}

transformed parameters {
    real<lower=0> length_scale = exp(log_length_scale);
    matrix[num_rows_padded, num_cols_padded 
        gp_periodic_matern_cov_rfft2(1.5, num_rows_padded, num_cols_padded,
        sigma, [length_scale, length_scale]',
        [num_rows_padded, num_cols_padded]');
    matrix[num_rows_padded, num_cols_padded] f = gp_inv_rfft2(
        z, rep_matrix(mu, num_rows_padded, num_cols_padded), rfft2_cov);
}

model {
    to_vector(z) ~ std_normal();
    mu ~ student_t(2, 0, 1);
    sigma ~ student_t(2, 0, 1);
    kappa ~ student_t(2, 0, 1);
    for (i in 1:num_rows) {
        for (j in 1:num_cols) {
            if (frequency[i, j] >= 0) {
                frequency[i, j] ~ neg_binomial_2(exp(f[i, j]), 1 / kappa);
            }
        }
    }
    // We use an implicit uniform prior on `log_length_scale`.
}
\end{Code}

We fitted the model to 80\% of the quadrants chosen uniformly at random using the below \proglang{Python} code, withholding the remainder for evaluation.
\begin{CodeInput}
>>> from gptools.stan import get_include
>>> import numpy as np

>>> # Load tree frequency matrix, define padding, apply training mask.
>>> frequency = np.loadtxt("tachve.csv", delimiter=",", dtype=int)
>>> num_rows, num_cols = frequency.shape
>>> padding = 10
>>> train_mask = np.random.binomial(1, 0.8, frequency.shape)
>>> data = {
...     "num_rows": num_rows,
...     "num_rows_padded": num_rows + padding,
...     "num_cols": num_cols,
...     "num_cols_padded": num_cols + padding,
...     "frequency": np.where(train_mask, frequency, -1),
... }

>>> # Compile model and fit it.
>>> model = compile_model(
...     stan_file="trees.stan",
...     stanc_options={"include-paths": get_include()},
... )
>>> fit = model.sample(data)
>>> print(fit.diagnose())
\end{CodeInput}
\begin{CodeOutput}
Processing csv files: ...

Checking sampler transitions treedepth.
Treedepth satisfactory for all transitions.

Checking sampler transitions for divergences.
No divergent transitions found.

Checking E-BFMI - sampler transitions HMC potential energy.
E-BFMI satisfactory.

Effective sample size satisfactory.

Split R-hat values satisfactory all parameters.

Processing complete, no problems detected.
\end{CodeOutput}

Despite the noisy, masked observations, the model was able to learn a smooth estimate of the density of trees, as shown in panel~(c). The posterior median of the correlation length of 60~m was well below the padding of 200~m introduced to attenuate the effect of periodic boundary conditions, as shown in panel~(b). We used a scaled mean-squared error (SMSE) to evaluate the model on the held-out data and compare it with the simpler approach of smoothing the data with a two-dimensional Gaussian filter. The SMSE is
\[
S\parenth{\vec{y},\hat{\vec{y}}=\exp\hat{\vec{f}}} = \frac{1}{m}\sum_{j=1}^m \frac{\parenth{y_i-\exp \hat f_i}^2}{\max\parenth{y_i,1}},
\]
where the sum is over $m$ test points and $\hat{\vec{f}}$ is the posterior median of the latent GP. We divided each term by the observed count $y_i$ (or one if the count was zero) to ensure the measure was not dominated by large counts because the sampling variance of a Poisson count process (without overdispersion) is equal to its mean.

A simple method to estimate the number of trees in held-out quadrants is to apply a Gaussian filter to the data and compare the two methods. The Gaussian filter estimate is
\[
    \hat{\vec{y}}_\lambda = \frac{\vec{g}_\lambda\ast\parenth{\vec{b}\circ\vec{y}}}{\vec{g}_\lambda\ast\vec{b}},
\]
where $\ast$ denotes convolution, $\circ$ denotes the elementwise product, $\vec{g}_\lambda$ is a Gaussian filter with smoothing scale $\lambda$, and $\vec{b}$ is the binary mask indicating which data are available for training. Gaussian filters ``blur'' the data locally such that adjacent elements of the smoothed signal can inform one another \citep{Lindeberg1990}. For large $\lambda$, estimates are approximated by the sample mean, and, for small $\lambda$, they are dominated by local noise. Panel~(d) of \cref{fig:trees} shows the SMSE for the Gaussian filter as a function of smoothing scale and the SMSE achieved by the GP model. The latter achieves a lower SMSE than the former for all smoothing scales, illustrating the utility of GPs for modeling spatial effects. Unlike in \cref{sec:tube}, it was not possible to use the posterior predictive distribution for evaluation because the Gaussian filter is not a generative model.

\section[Discussion]{Discussion\label{sec:discussion}}

We implemented two popular approaches for scaling GPs to larger datasets in \proglang{Stan}: The sparse approximation with structured dependencies discussed in \cref{sec:graph} and the exact Fourier approach in \cref{sec:fourier} which is applicable to data on a grid. For centered parameterizations, the likelihood can be evaluated or approximated directly. For non-centered parameterizations, we sample standard normal random variables $\vec{z}$ and use the inverse transform to obtain a GP sample $\vec{f}=\phi^{-1}\parenth{\vec{z}}$.

Given different parameterizations and approaches, which should be used in practice? As discussed in \cref{sec:benchmark}, a non-centered parameterization is appropriate if the data are weak, and a centered parameterization is preferable if the data are strong. Choosing the right parameterization ensures parameters are relatively uncorrelated under the posterior distribution which accelerates inference. For variational mean-field approximations, choosing the right parameterization is even more important: It affects the quality of the approximation, as discussed in \cref{sec:benchmark}. Most variational approaches use a centered parameterization \citep{Hensman2013,Wu2022}, and their approximations may be improved by considering non-centered parameterizations. If the data are very strong, the benefits of GPs may be outweighed by their complexity because the likelihood dominates the GP prior. If in doubt, we suggest using a non-centered parameterization, as we have done in \cref{sec:illustration}, because GPs are typically employed when the data are not sufficiently informative for simpler approaches to succeed.

\begin{figure}
    \centering
    \includegraphics{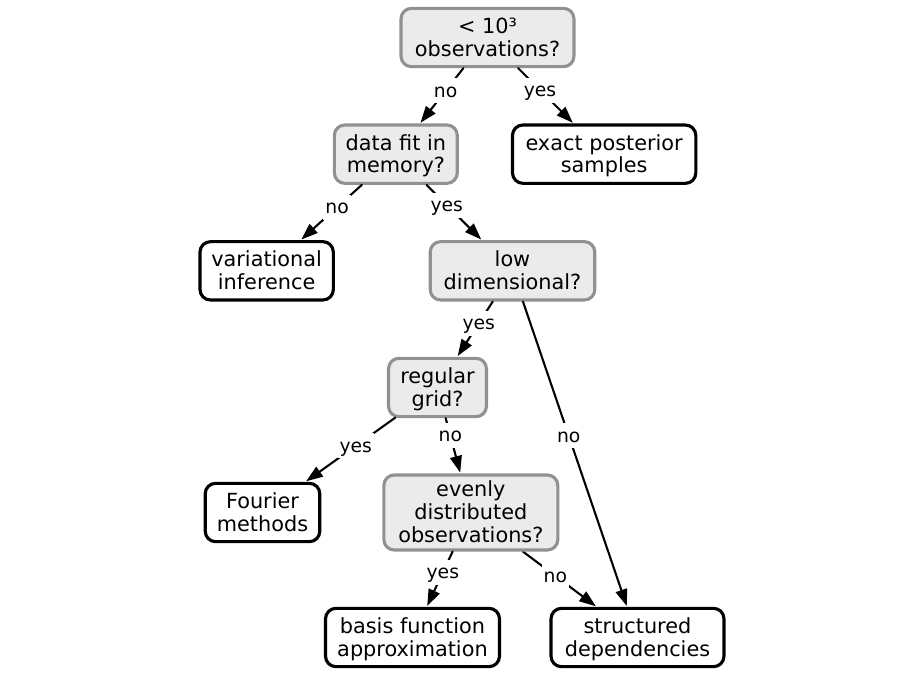}
    \caption{The optimal inference approach and package depends on the data.}
    \label{fig:decision-tree}
\end{figure}

Choosing an appropriate inference approach and between the many packages discussed in \cref{sec:introduction} depends on the data at hand. Here, we only consider general purpose packages that can handle arbitrary likelihoods and facilitate the use of GPs as parts of larger models, as shown in \cref{fig:decision-tree}. At one extreme, posterior samples can be obtained with most inference frameworks if the data are small. At the other, if the data do not fit in memory, evaluating the likelihood repeatedly as part of a Monte Carlo sampler is not feasible---even if the likelihood can be approximated. ADVI (e.g., using \pkg{GPyTorch} and \pkg{Pyro}) is a viable approach although at the cost of considering a narrower set of posteriors \citep{Kucukelbir2017}. If the data are low-dimensional and form a regular grid, the Fourier methods in \cref{sec:fourier} are suitable. Basis function approximations (currently implemented in the \proglang{R} package \pkg{brms} which builds on \proglang{Stan}) may be appropriate for low-dimensional data with irregular spacing. The implementation represents the GP as a linear superposition of eigenfunctions of the Laplace operator with Dirichlet boundary conditions \citep{Riutort-Mayol2022}. The approach is similar to the Fourier methods presented here because Fourier modes are eigenfunctions of the Laplace operator although with periodic boundary conditions. The unique advantage of Fourier methods is that the likelihood can be evaluated exactly in $\mathcal{O}\parenth{n\log n}$ if observations form a grid. In higher dimensional spaces, sparse approximations using structured dependencies can approximate the posterior, as discussed in \cref{sec:graph}. A meaningful performance comparison between packages is challenging because they seek to answer different questions (e.g., maximum marginal likelihood estimation or full posterior inference), employ different methods to answer the same question (e.g., ADVI, non-adaptive leapfrog sampler, Gibbs sampling, or adaptive No-U-Turn sampler), and use different programming languages with vastly different performance (e.g., compiled \proglang{C++} or interpreted \proglang{Python}).

Padding may be required to attenuate the effect of periodic boundary conditions inherent to the fast Fourier transform. The necessary amount of padding depends on the kernel. We have found one to two correlation lengths $\ell$ to be sufficient for squared exponential and Mat\'ern~$\sfrac{3}{2}$ kernels (see \cref{app:padding} for details). However, the correlation length is often not known a priori, and finding the ``right'' amount of padding that appropriately balances performance and the need for non-periodic boundary conditions may be an iterative process. For example, we can start with a small amount of padding and increase it until the posterior stabilizes.

Fourier methods may also be appropriate if the density of observation points is relatively homogeneous. In particular, we may consider a latent GP $\vec{g}$ on a grid and use it to predict the GP of interest $\vec{f}$ at each observation point, i.e.,
\[
    \proba{\vec{f}\mid\vec{g}}=\prod_{j=1}^n \proba{f_j\mid \vec{g}}.
\]
This method reduces the computational cost because elements of $\vec{f}$ are conditionally independent given $\vec{g}$ at the regularly spaced ``inducing points'' \citep{Hensman2013}.

We hope that our library and the illustrations in \cref{sec:illustration} will accelerate the development of models employing GPs in \proglang{Stan}. Integrating GP approximations with \proglang{Stan}'s ecosystem, rather than developing a bespoke GP library, will allow practitioners to leverage the framework's flexibility and the shared knowledge of the engaged \proglang{Stan} community.

\section*{Computational details}

The results in this paper were obtained using \proglang{Python}~3.10.13, \pkg{cmdstanpy}~1.1.0, and \pkg{cmdstan}~2.33.0. All experiments were run on a single core of a 2020~MacBook Pro with an Apple Silicon M1 chip and 16~GB of RAM.

\section*{Acknowledgments}

We thank Philip Greengard, Mike Lawrence, and Aki Vehtari for comments on the manuscript and Brian Ward for answering numerous questions about \pkg{cmdstanpy}.

\bibliography{main}

\newpage
\begin{appendix}

\section[Padding]{Effect of padding for Fourier methods\label{app:padding}}

We considered a simulation study to examine the effect of periodic boundary conditions inherent to Fourier methods and assess the amount of padding required to balance model misspecification and performance concerns in two steps.

\begin{figure}
    \centering
    \includegraphics{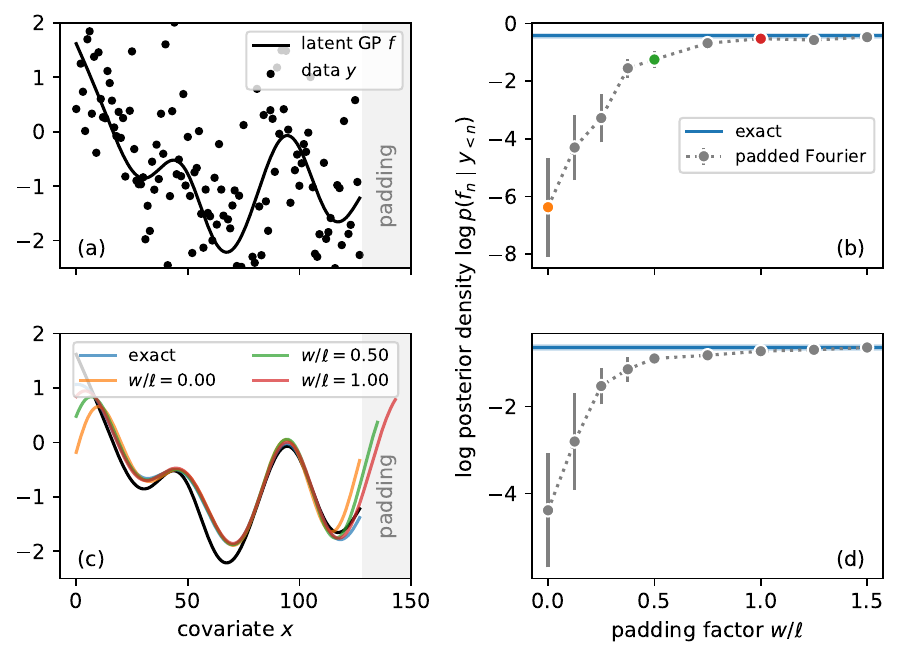}
    \caption{A small amount of padding is sufficient to attenuate the effect of periodic boundary conditions. Panel~(a) shows a realization of the benchmark model in \cref{eq:benchmark-centered}, and panel~(c) shows the true latent GP $f$ together with posterior means for different models, including the true generative model and Fourier methods with varying amounts of padding $w$. The effect of periodic boundary conditions is stark for the fit without padding, but it disappears quickly with increasing padding. Panels~(b) and~(d) show the log posterior density evaluated at the true $f_n$, which corresponds to the held-out data point $y_n$, as a function of padding for squared exponential and Mat\'ern~$\sfrac{3}{2}$ kernels, respectively. The horizontal line corresponds to the log posterior density obtained using the true generative model. Padding with one length scale $\ell$ is sufficient to approximate the true model well.}
    \label{fig:padding}
\end{figure}

First, we generated $m=100$ synthetic datasets each comprising $n=128$ observations on an integer grid according to the benchmark model in \cref{eq:benchmark-centered} with marginal kernel scale $\sigma=1$ and observation noise $\kappa=1$. We used a correlation length $\ell=16$ large enough for periodic boundary conditions to have an effect on the inference. For each dataset, we fitted the standard non-centered model, i.e., the true generative model, and Fourier-based Gaussian processes with varying amounts of padding $w$. We evaluate both models by holding out the last data point $y_n$, which should be most severely affected by periodic boundary conditions, and approximating the log posterior density $\log\proba{f_n\mid y_{<n}}$ of the corresponding element of the latent GP using a Gaussian kernel density estimator \citep[Chapter~2.5.1]{Bishop2006}. Even a small amount of padding, such as one correlation length, is sufficient to attenuate the effect of periodic boundary conditions, as shown in panels~(b) and~(d) of \cref{fig:padding} for squared exponential and Mat\'ern~$\sfrac{3}{2}$ kernels, respectively.

Simulation-based calibration is a technique to validate a Bayesian inference pipeline~\citep{Talts2020}. For synthetic data generated from the model, the rank of the true parameter value among posterior samples should have a uniform distribution. For each combination of the different paddings and two kernels, we evaluated the rank across $m$ synthetic datasets and evaluated the $p$~value of the Kolmogorov-Smirnov test by comparing with a discrete uniform reference distribution. If no padding is used, the null hypothesis that the ranks are uniform can be confidently rejected ($p$~value $<10^{-3}$), but the ranks are not inconsistent with a uniform distribution for $w>\ell$ ($p$~value $>0.3$ at $w=\ell$ in our simulations).

\section[Kernels]{Kernels in the real and Fourier domains\label{app:kernel-expressions}}

\subsection[Squared exponential kernel]{Squared exponential kernel}
The non-periodic squared exponential kernel is defined as
\[
k\parenth{x,x'}=\sigma^2\exp\parenth{-\frac{\parenth{x-x'}^2}{2\ell^2}},
\]
where $\sigma$ is the marginal scale and $\ell$ is the correlation length. Its discrete power spectrum on a periodic domain of size $L$ is
\[
\tilde k_\xi= \frac{\sqrt{2\pi} n \sigma ^ 2 \ell}{L}  \exp\parenth{-2\parenth{\frac{\pi\xi\ell}{L}} ^ 2},
\]
where $n$ is the number of grid points and $\xi\in\bracket{0..n-1}$ is the discrete frequency.

\subsection[Mat\'ern kernel]{Mat\'ern kernel}
The non-periodic Mat\'ern kernel is defined as
\begin{align*}
k_\nu\parenth{x,x'} &= \sigma^2\frac{2^{1-\nu}}{\Gamma(\nu)}\zeta^\nu K_\nu\parenth{\zeta},\\
\text{where }\zeta&=\frac{\sqrt{2\nu}\abs{x-x'}}{\ell}
\end{align*}
is a rescaled distance, $\nu$ is a smoothness parameter, $\Gamma$ denotes the gamma function, $\abs{x-x'}$ is the Euclidean distance between $x$ and $x'$, and $K_\nu$ denotes the modified Bessel function of the second kind. For $\nu=\sfrac{3}{2}$ and $\nu=\sfrac{5}{2}$, the kernel simplifies to
\begin{align*}
k_{\sfrac{3}{2}}&=\sigma^2\parenth{1+\frac{\sqrt 3 \abs{x-x'}}{\ell}}\exp\parenth{-\frac{\sqrt 3 \abs{x-x'}}{\ell}}\\
k_{\sfrac{5}{2}}&=\sigma^2\parenth{1+\frac{\sqrt 5 \abs{x-x'}}{\ell} + \frac{5\abs{x-x'}^2}{3\ell^2}}\exp\parenth{-\frac{\sqrt 5 \abs{x-x'}}{\ell}}.
\end{align*}
It reduces to the Laplace kernel for $\nu=1/2$. Its discrete power spectrum on a periodic domain of size $L$ is
\[
\tilde k_\xi = \sigma^2 \frac{n\ell}{L} \parenth{\frac{2 \pi}{\nu}} ^ {p / 2}
        \frac{\Gamma\parenth{\nu + \frac{p}{2}}}{\Gamma\parenth{\nu}}
        \parenth{1 + \frac{2 (\pi \ell \xi) ^ 2}{\nu L^2}} ^ {-(\nu + p / 2)},
\]
where $p$~is the dimensionality of the space, $n$ is the number of grid points, and $\xi\in\bracket{0..n-1}$ is the discrete frequency.

\end{appendix}

\end{document}